\documentclass[conference, 10pt]{IEEEtran}
\IEEEoverridecommandlockouts
\usepackage{tabularx}
\usepackage{cite}
\usepackage{amsmath,amssymb,amsfonts}
\usepackage{algorithmic}
\usepackage{graphicx}
\usepackage{textcomp}
\usepackage{comment}
\usepackage{multirow}
\usepackage{booktabs}
\usepackage[linesnumbered,lined,boxed,commentsnumbered,ruled,longend, noend]{algorithm2e}
\SetKwInput{KwInput}{Input}                
\SetKwInput{KwOutput}{Output}
\usepackage{fancyhdr}
\usepackage{array}
\usepackage{makecell}

\usepackage[dvipsnames]{xcolor}

\begin{document}

\title{A Novel Buffered Federated Learning Framework for Privacy-Driven Anomaly Detection in IIoT\\

}
\vspace{-1cm}
\author{
    \IEEEauthorblockN{Samira Kamali Poorazad, Chafika Benza\"{i}d, and Tarik Taleb}
    \IEEEauthorblockA{
        Oulu University, Finland; Oulu University, Finland; Ruhr University Bochum (RUB), Germany\\
        Emails: samira.kamalipoorazad@oulu.fi, chafika.benzaid@oulu.fi, tarik.taleb@rub.de
    }

\vspace{-1.1cm}
}

\maketitle

\fancypagestyle{mahmood}{%
   \fancyhf{} 
   \renewcommand{\headrulewidth}{0pt}
   \fancyhead[C]{Reprinting or republishing this material for the purpose of advertising or promotion, reselling or redistributing to servers or lists, or using any copyrighted component in other works must adhere to IEEE policy. The paper has been accepted for publication at \textbf{GlobeCom 2024}, and DOI will be provided as soon as possible.}
}

\thispagestyle{mahmood} 

\begin{abstract}

Industrial Internet of Things (IIoT) is highly sensitive to data privacy and cybersecurity threats. Federated Learning (FL) has emerged as a solution for preserving privacy, enabling private data to remain on local IIoT clients while cooperatively training models to detect network anomalies. However, both synchronous and asynchronous FL architectures exhibit limitations, particularly when dealing with clients with varying speeds due to data heterogeneity and resource constraints. Synchronous architecture suffers from straggler effects, while asynchronous methods encounter communication bottlenecks. Additionally, FL models are prone to adversarial inference attacks aimed at disclosing private training data. To address these challenges, we propose a Buffered FL (BFL) framework empowered by homomorphic encryption for anomaly detection in heterogeneous IIoT environments. BFL utilizes a novel weighted average time approach to mitigate both straggler effects and communication bottlenecks, ensuring fairness between clients with varying processing speeds through collaboration with a buffer-based server. The performance results, derived from two datasets, show the superiority of BFL compared to state-of-the-art FL methods, demonstrating improved accuracy and convergence speed while enhancing privacy preservation.

\end{abstract}


\section{Introduction}
Industrial Internet of Things (IIoT) is a form of Internet of Things (IoT) implemented in manufacturing and industry to automate processes and produce more effective and appropriate products \cite{samira}. IIoT offers a multitude of advantages over traditional Supervisory Control and Data Acquisition (SCASA) systems, including productivity, scalability, and data analysis \cite{samira}. Neverthless, 
the increase in connected devices and the lack of security design in older control systems make factories vulnerable to cyber-attacks such as denial of service (DoS) \cite{samira}. In addition to security challenges, data privacy is another critical concern in IIoT environments. This is attributed to the massive amount of data generated by IIoT smart devices, which may encompass sensitive information that service providers are reluctant to disclose to third parties  
\cite{samira}. As a result, an imperative arises 
for a solution to detect cyber-attacks while maintaining data privacy in IIoT domains.

Many IIoT contexts have employed centralized machine learning-based anomaly detection schemes as a solution \cite{CommEfficient}. In most cases, centralized schemes lack the flexibility to adapt to different scenarios, particularly with regard to privacy concerns \cite{smartbuildings}. This limitation arises from the inherent need of centralized methods to send all IIoT data 
to a single location for processing. This makes centralized approaches time-consuming (causing inefficiency in communication) and compromises the privacy of sensitive industrial processes \cite{DÏoT}. 
To \textit{preserve privacy} and improve communication efficiency, federated learning (FL) emerges as an appropriate distributed machine learning technique for IIoT environments \cite{smartbuildings}. FL allows clients to train a global model collaboratively by transmitting parameters and local models to a server instead of sharing raw data \cite{DÏoT}, \cite{farhoudi}. This reduces training time and addresses privacy concerns. FL frameworks support synchronous and asynchronous modes of communication.

Synchronous FL (SFL) \cite{FEDAVG} requires all clients to submit their local models to the server for aggregation. However, problems may arise when clients train at different speeds due to resource differences, such as differences in memory capacity or heterogeneous data. In this situation, the aggregation server must wait for the slowest client to transmit its local model, resulting in slow convergence speeds and delays. This delay, known as the  \textit{straggler effect}, occurs due to resource differences, causing certain clients to lag behind. Therefore, SFL is not suitable for real-time applications. To avoid Straggler effects, asynchronous FL (AFL) \cite{asynchFL} was introduced, where aggregation occurs immediately after a client trains its model, without waiting for the slowest client. Nevertheless, most asynchronous approaches experience \textit{communication bottlenecks} because clients can communicate independently with the aggregation server. 

To trade off SFL and AFL, buffered base solutions have been proposed. As an example, the buffered FL approach (Fed-Buff) \cite{fedbuff} employs a K-size buffer at the server, processing client updates only after receiving K updates. A notable limitation of Fed-Buff is its inability to adapt to speed differences among heterogeneous clients. As a result, the buffer is overwhelmed by faster clients, introducing training bias.
Moreover, without considering the differences in data and computational speed among clients, the accuracy of the global model could suffer as the local models of faster clients might be less accurate. Consequently,  when leveraging different methods of FL within an IIoT heterogeneous environment, it is necessary to take into account the \textit{tradeoffs between model convergence speed, model accuracy, and balance between clients with varying speeds}.

Furthermore, although FL method provides privacy preservation compared to traditional centralized methods, it is still vulnerable to privacy attacks involving model poisoning and model inference \cite{AIB5G}. Attackers may eavesdrop on the communication channel or compromise the server to access model updates, resulting in data leakage \cite{AIB5G}, \cite{somayeh}. Thus, it is essential to \textit{enhance the privacy of FL} in addition to focusing on privacy concerns in IIoT through FL. Homomorphic encryption (HE) can be used to enhance FL privacy by aggregating encrypted model updates \cite{AIB5G}. In contrast to some secure FL schemes, such as differential privacy (DP)-based FL, which introduce inevitable accuracy losses due to added noise, FL with HE preserves model accuracy while protecting data privacy \cite{FusionFL}. 


The challenges previously highlighted motivate us to develop a Buffered FL (BFL)-based anomaly detection model for IIoT.  This model effectively addresses three critical challenges from a new perspective: stragglers, communication bottlenecks, and privacy-preservation. BFL achieves this 
by combining synchronous and buffered FL concepts in a novel manner, despite heterogeneous client speeds. A global model in BFL is synchronously transmitted only to clients who have submitted their local parameters to the server's buffer within a specified time. Based on the training times of each client, the specific time is calculated using a novel weighted average time method. The proposed weighted average time method is designed to ensure fairness between fast and slow clients, minimizing the straggler effects, communication bottlenecks, and training bias. Moreover, HE-based secure communication is implemented to enhance model parameter privacy. The main contributions of this work are as follows:

\begin{enumerate}
  \item Proposing a new anomaly detection framework that leverages a deep learning (DL)-based FL approach and HE to empower privacy-enhanced detection of cyber threats in industrial cyber-physical systems (CPS). 
  \item Developing BFL, a novel FL framework that combines the potential of synchronous and buffered FL approaches to address straggler and communication issues. 
  \item Designing a novel weighted average time method to balance a trade-off between fast and slow clients.
  \item Utilizing two distinct datasets to demonstrate the generalization capabilities of BFL.  
\end{enumerate}

This article is organized as follows. Section ~\ref{sec:sec2} categorizes some related articles. Section \ref{sec method} introduces the proposed method, system model, and attack model. Section \ref{eval} presents implementation details and discusses the evaluation results. The article concludes in Section \ref{conc}.

\section{Related Work}\label{sec:sec2}
This section provides a categorized review of FL-based intrusion/anomaly detection schemes for IIoT environments.

\subsection{FL-based Intrusion/Anomaly Detection}

\subsubsection{SFL approaches} The authors of \cite{CommEfficient} proposed a communication efficient-FL framework based on Convolutional Neural Network (CNN)-Long Short-Term Memory (LSTM) to reduce communication costs by using gradient compression and local computations. In order to identify the best gradients, the Top-k algorithm is used. In \cite{smartbuildings}, a FL approach based on LSTM is proposed for detecting anomalies in energy consumption in smart buildings. Authors in \cite{DÏoT} proposed an anomaly detection system based on federated self-learning to detect malicious devices in IoT. The system uses Gated Recurrent Units (GRUs) to classify data based on thresholds. Additionally, the self-learning mechanism improves the detection performance of the global model as the IoT environment changes.

The study in \cite{ICSDetectAttack} aims at detecting anomalies in industrial control systems (ICS), leveraging a Variational Autoencoder-LSTM model to capture temporal dependencies and complex patterns. 
 The work in \cite{ensembleFL} presents IIoT cyber-threat hunting model that uses a One-Class Support Vector Machine (OCSVM), Isolation Forests (IF), and Stacked Autoencoders (SAE). A combination of OCSVM and IF is used to detect potentially malicious data points. The SAE helps identify patterns and relationships in data by extracting features.

In addition to the lack of privacy-preserving methods, the discussed synchronous methods are particularly susceptible to straggler effects when more clients are involved. The straggler effect results in delays and slow convergence in real-time industrial domains.

\subsubsection{AFL approaches}
 In \cite{AFLElephantSDN}, an AFL is proposed for Software Defined Networking systems with distributed control to improve efficiency. Models are trained by local controllers and are uploaded to root controllers asynchronously. Local models are aggregated by root controllers. In \cite{CSAFL}, a spectral clustering method based on the latency and direction of model updates is proposed to address the issue of model staleness caused by asynchronous updates. The scheme also improved test accuracy and convergence speed in non-independent and identically distributed (non-IID) datasets. In \cite{digital_twin_AFL}, the authors proposed an AFL-based digital twin architecture for IIoT applications to minimize straggler effects. Results showed faster convergence and higher learning rates with the suggested model. In \cite{SAFA}, the authors propose a Semi-Asynchronous Federated Averaging (SAFA) to solve low round efficiency and poor convergence by using a novel aggregation algorithm with a cache structure.

Investigated 
asynchronous methods require more frequent communication between clients and server, resulting in higher communication costs. Convergence is also slowed by frequent aggregation on the server.

\subsection{Privacy-enhancing FL}

Although FL methodology offers privacy-preserving benefits, there are still privacy concerns associated with it. Various privacy-enhancing technologies can be used to enhance the effectiveness of FL.


In \cite{HECovid-19}, FL-CNN is used to minimize sensitive information sharing, and HE is leveraged to enhance privacy. In \cite{BlockchainFL}, authors introduced a novel distributed FL scheme in an IIoT scenario based on K-means, distributed random forest, and AdaBoost algorithms with the incorporation of DP and HE techniques to enhance data privacy and provide safe data sharing among IIoT devices. 
The work in \cite{reliableFL} proposes an approach to enhance privacy preservation of FL using HE. It addresses common issues such as privacy breaches, communication overheads, and lack of accountability.
In \cite{DeepFed}, a FL based on CNN and GRU was employed to improve ICS intrusion detection accuracy by learning both spatial and temporal features of network traffic data. HE is utilized in this method to enhance privacy.

Various privacy-enhancing techniques, such as DP and HE, have been successfully integrated into FL approaches to enhance data privacy in industrial scenarios. While DP is effective in protecting client privacy during aggregation, HE is preferred in certain cases to enhance the privacy of model updates and ensure data/parameter confidentiality. In this work, our objective is to improve the privacy of model updates using HE, as DP can introduce noise affecting model performance, such as utility.

\section{Methodology}\label{sec method}
This section introduces the system and attack models as well as the proposed methodology for empowering privacy-driven FL-based anomaly detection in IIoT.

\subsection{System Model}
\begin{figure}[htbp]
\centerline{\includegraphics[width=3.5in]{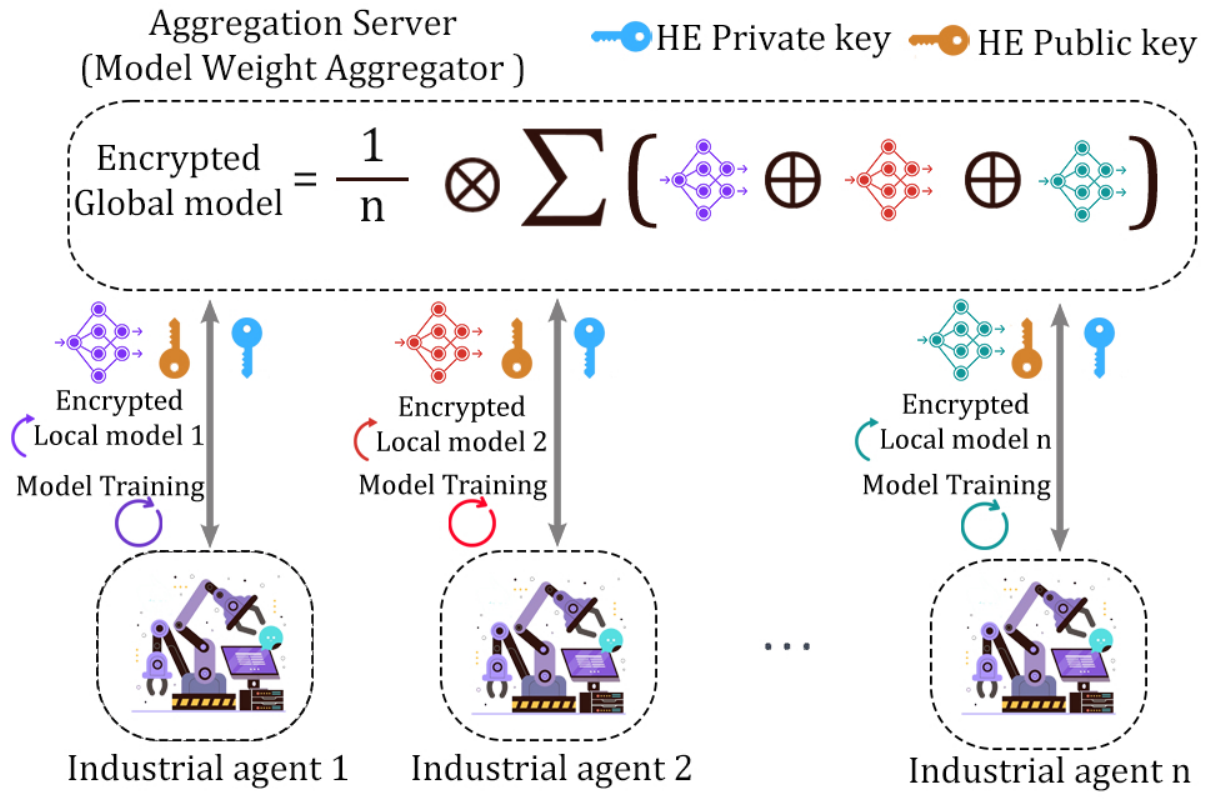}}
\caption{High-level architecture for privacy-preserving BFL.}
\label{figure:system_model}
\vspace{-0.5cm}
\end{figure}

Fig. \ref{figure:system_model} illustrates a high-level system overview of the proposed privacy-preserving FL training method. It consists of industrial agents and an aggregation server. Industrial agents represent owners of CPSs. A third authority generates public and private keys first. Industrial agents construct a local DL model using their own industrial CPS data, encrypt the parameters of the model using the public key, and send them to the aggregation server. The aggregation server aggregates the encrypted model parameters from each industrial agent. The aggregated encrypted parameters are then sent back to the industrial agents. The industrial agents update their DL models by decrypting the aggregated parameters with the private key. This process is repeated until the model is converged.
\setlength{\textfloatsep}{0pt}





\subsection{Attack model}
The threat model includes cyber threats to industrial CPSs and the BFL framework. To provide a comprehensive understanding of the risks faced by industrial CPSs and the BFL framework, this work examines the following cyber threats:

\begin{enumerate}
 \item Cyber Threats Against Industrial CPSs: We examine
 various cyber threats to industrial CPS remote processes, including response injection attacks, command injection attacks, reconnaissance attacks, and DoS attacks.  
 Response injection attacks involve falsified response messages to querying entities, while command injection attacks introduce false commands into the industrial control systems. Reconnaissance attacks aim to gather information about industrial CPSs, and DoS attacks overload a target system, causing disruptions in industrial CPSs.
    
    
    
\item Cyber Threats Against BFL Framework: External attackers or malicious eavesdroppers may also intercept communication links. This study considers eavesdropping on the model parameters. Data resources can be analyzed through the parameters of the anomaly detection model in this attack.
\end{enumerate}

\subsection{Proposed Methodology} \label{bb}
A novel FL framework and a HE-based secure communication protocol are used to network multiple industrial CPS owners to develop a DL anomaly detection model. We use a partial HE; a type of HE that permits both the addition and multiplication of ciphertexts.  The rationale behind this choice is rooted in the superior speed of partial HE compared to fully HE. It is also motivated by the fact that partial HE is,
in most cases, 
sufficient for homomorphic calculations, especially for FLs that only require addition and multiplication. 

The BFL framework employs the SFL strategy to avoid communication bottlenecks. In this manner, the updated model will not be sent individually to each industrial agent, such as AFL. A buffer-based server is also used to prevent stragglers. Thus, BFL server does not wait for all agents to send their local models before aggregating. Rather, BFL server waits until a specific time for agents to send their local models. To calculate the specific time, avoid overloading the buffer-based server, and avoid training bias, the weighted average time method is used during the first iteration of BFL. The weighted average time method determines the specific time based on the training time spent by each agent. 
Algorithms \ref{alg:detection_prevention} and \ref{alg:weighted_time} summarize the detailed training procedure.

The trust authority first generates a public key and a private key. The aggregator server selects initial parameters. In the first iteration of BFL, agents receive initial parameter values from the aggregator server, they train DL-based anomaly detection models locally with their own data (line 7 of Algorithm \ref{alg:detection_prevention}). As each agent trains a local model, the model parameters are encrypted using the public key (line 8 of Algorithm \ref{alg:detection_prevention}). Afterward, the encrypted parameters and training time for each agent are uploaded to the aggregator. After the aggregator server receives the training times and local encrypted parameters of the agents, encrypted global parameters and weighted average time are calculated (lines 13 and 15 of Algorithm \ref{alg:detection_prevention}). As a first step in calculating weighted average time, the aggregator server sorts the received training times in descending order (line 1 of Algorithm \ref{alg:weighted_time}). Each agent's initial weight is derived by inverting training time (line 3 of Algorithm \ref{alg:weighted_time}). However, to assign higher weights to slower agents and lower weights to faster agents, the Initial weights must be reversed to achieve the final weights (line 4 of Algorithm \ref{alg:weighted_time}).

\setlength{\intextsep}{4pt}
\begin{algorithm}[]
\footnotesize
\caption{Detection Algorithm}
\label{alg:detection_prevention}
\KwInput{\begin{tabbing}
    \=$T$: Number of Iterations; $N$: Number of Clients (Agents);\\
    \> $P_{key}$: Public key, $S_{key}$: Private key;\\
    \> $\oplus$: HE-based addition; $\otimes$: HE-based multiplication,\\ $S_m$: A\ list\ of\ selected\ clients
    \end{tabbing}



}
\KwOutput{Desired performance (e.g., accuracy)}

{
     $m = Initial\_parameters()$ \\
     $M_{agg} = 0$  \  \# Encrypted global parameters \\
    \For{$t=1$ to $T$}
    {
        \If{$t = 1$}
        {
            \For{each client $k=1$ to $N$}
            {
                $t_s = \text{Start\_Time()}$ \\
                $[Local\_m]^{k} = \text{$m.train()$}$ \\
                $[m_{enc}]^k = \text{$Encrypt([Local\_m]^k,P_{key})$}$ \\
                $t_e =  \text{End\_Time()}$\\
                $[T_{\text{client}}]^{k} = t_e - t_s$
            }
            \# Encrypted aggregation on Server\\
            \For{$j$ $\in [m_{enc}]$}
            {
                $[M_{\text{agg}}] = \text{$(j \oplus M_{\text{agg}})$} \otimes [N^{-1}]$ \\
            }         
            \# Calculating weighted average time\\
            $T_{\text{W\_Avg}} = \text{Weighted\_Average\_Time}([T_{\text{client}}])$

            \For{$i$ $\in [T_{\text{client}}]$}
            {
                \If{$i \leq T_{\text{W\_Avg}}$}{
                    $[S_m] = i$
                }
            }

        }
        \Else
        {
            \For{each client $k \in [S_m]$}
            {
                    $[m_{dec}]^k =\text{$Decrypt([M_{agg}],S_{key})$}$\\
                    $[Local\_m]^k = \text{$m_{dec}.train()$}$ \\
                    $[m_{enc}]^k = \text{$Encrypt([Local\_m]^k,P_{key})$}$
            }

            \# Encrypted aggregation on Server\\
            \For{$j \in\ [m_{enc}]$}
            {
                $[M_{\text{agg}}] = \text{$(j \oplus M_{\text{agg}})$} \otimes [N^{-1}]$ \\
            }
        }
    }
}
\end{algorithm}
\vspace{-1em} 
\setlength{\intextsep}{0.1pt}
\normalsize

\begin{algorithm}[]
\footnotesize
\caption{Weighted\_Average\_Time}
\label{alg:weighted_time}
\KwInput{$Trainig\ Time\ of\ each\ client\ (Agent)$ = $T_{client}^{1},T_{client}^{2},..., T_{client}^{N}$}

\KwOutput{Weighted Average Time}

{
    $Sort\ {T_{client}}^{1},{T_{client}}^{2},..., {T_{client}}^{N}\ in\ descending\ order$
    \For{each $i=1$ to $n$}{
    
        \# Initial Weight = [$W_1, W_2, ..., W_N$]\\
        
        $[\text{Initial\_Weight}]^{i} = \frac{1}{{\text{$T_{client}$}}^{i}}$
        
    }

    \# Reverse Initial weight
    $[\text{Final\_weight}] = [W_N, W_{N-1}, ..., W_1]$

        $W_f = \sum_{j=1}^{N}{\text{Final\_weight}}^{j}$

        $T_{fw} = \sum_{j=1}^{N}{\text{Time\_training}}^{j} \cdot \text{Final\_weight}^{j}$
    
    $T_{\text{W\_Avg}} = \frac{T_{fw}}{W_f}$
}
\end{algorithm}
\setlength{\intextsep}{0.11pt}
\normalsize

Finally, the average weighted time can be obtained by multiplying each agent's training time by its corresponding final weight over the sum of all final weights (line 7 of Algorithm \ref{alg:weighted_time}). The encrypted global parameters are then sent back to the agents. Agents obtain updated model parameters by decrypting the encrypted parameters model with a private key. After that, the DL model parameters are updated. For the second iteration of BFL to convergence, only agents with training times less than the calculated weighted average time will be selected (lines 19 to 26 of Algorithm \ref{alg:detection_prevention}).

\section{Experiment Setting and Evaluation}\label{eval}
\vspace{-0.05cm}
\subsection{Experimental Settings}
We evaluated BFL using two different datasets:
(i) Gas pipeline dataset \cite{gaspip}, which consists of 18 features related to Modbus protocol in 8 classes including normal, DoS, Naive/Complex Malicious Response Injection, Malicious State/Parameter Command Injection, Malicious Function Code Injection, and Reconnaissance; and (ii) the WUSTL-IIoT Dataset \cite{wsul-IIoT}, which consists of 41 features of real cyber-attacks related to network flow in 5 classes: DoS, Reconnaissance, Backdoor, Command injection, and Normal. 

BFL is compared to the three FL methods: FedBuff \cite{fedbuff}, SFL \cite{FEDAVG}, and AFL \cite{asynchFL}. We have implemented BFL in PyTorch and utilized the Paillier Framework for HE.

Both a CNN and a Multilayer Perceptron (MLP) were trained on two specified datasets. The CNN architecture consists of two 1D convolutional layers, two fully connected layers, one MaxPooling layer, and one AveragePooling layer. For both datasets, the first convolutional layer has 8 filters of size 3, and the second convolutional layer has 16 filters of size 3. The MaxPooling layer and AveragePooling layer, each has a kernel size of 2. The Rectified Linear Unit (ReLU) activation function is applied in the convolutional layers, and Softmax is used for multi-class classification. The MLP architecture applied to both datasets comprises three fully connected layers. For the Gas Pipeline dataset, the first layer maps the 18-dimensional input to a 54-dimensional space, the second layer reduces this to a 20-dimensional space, and the final layer maps it to 8 classes. For the WUSTL-IIoT dataset, the first layer maps the 41-dimensional input to a 9-dimensional space, the second layer also maps this to a 9-dimensional space, and the final layer maps it to 5 classes. Stochastic Gradient Descent (SGD) was selected as the optimization function. For the Gas pipeline dataset, a batch size of 64 was used, while a batch size of 1000 was used for the WUSTL-IIoT dataset. The learning rate and momentum were set at 0.01 and 0.8, respectively. These hyperparameters were determined through trial and error, with the combination of a 0.01 learning rate and 0.8 momentum yielding the best results. Due to space constraints, we present only the results of the model trained with SGD using these optimal hyperparameters. Datasets are divided into three parts: 80\% for training, 10\% for validation, and 10\% for testing. The data is then normalized using the MinMaxScaler, which scales the values between 0 and 1.

After defining the model with the aforementioned values and executing it, the model is evaluated using the accuracy metric, overall training time (millisecond (ms)), and convergence speed (i.e., the number of iterations required to achieve a target accuracy).
 
\subsection{Comparison of two DL algorithms on BFL}\label{AA}
As a first step, we compared the efficiency of BFL based on CNN and MLP algorithms in terms of accuracy, and overall training time to determine the best algorithm to use for the next steps. In order to compare them in terms of overall training time, we take into account 5 clients (industrial agents), and the target accuracy of 94.7\% for the Gas pipeline dataset and 99.8\% for WUSTL-IIoT dataset. Both accuracy targets are achieved when all data is available. The accuracy of both models is also compared based on 5 clients, two iterations, 10 local epochs, 0.01 learning rate, and stochastic gradient descent (SGD) optimization for both datasets. As shown in Table \ref{table:overalltimehfl} and Table \ref{table:acchfl}, MLP outperforms CNN in both datasets. The simplicity of the MLP architecture, with fewer parameters to learn, compared to CNN with more layers, such as pooling layers, allows it to converge faster to the target accuracy. Therefore, MLP is a more appropriate approach for real-time environments with tabular data. Consequently,  we select MLP  as the baseline algorithm for further comparisons in the next sections (\ref{xx} and \ref{yy}). 


\begin{table}[t!]
\caption{ BFL Overall training {time(ms)} with 5 clients.}
\vspace{-0.4cm}
\label{table:overalltimehfl}
\begin{center}
\begin{tabular}{|c|c|c|}
\hline

\thead{Modes} &  \thead{GAS Pipeline \\ Target accuracy: 94.7\%}  &  \thead{WUSTL-IIoT \\ Target accuracy: 99.8\%} \\ \hline
MLP & 1655.78 &  5965.07 \\ \hline
CNN& 1995.05 & 8465.18 \\ \hline
\end{tabular}
\end{center}
\vspace{-0.4cm}
\end{table}

\begin{table}[t!]
\caption{BFL accuracy evaluation with 5 clients and 2 iterations.}
\vspace{-0.4cm}
\label{table:acchfl}
\begin{center}
\begin{tabular}{|c|c|c|}
\hline

Modes & GAS Pipeline &  WUSTL-IIoT\\ \hline
MLP & 94.59{\%} & 99.70{\%} \\ \hline
CNN& 93.60\% & 98.49\% \\ \hline
\end{tabular}
\end{center}
\vspace{-0.1cm}
\end{table}

\subsection{Overall training time evaluation based on client numbers}\label{xx}

                                \begin{table}[t!]
                                \caption{{MLP-based BFL - Overall training time(ms) by increasing clients.}}
                                \vspace{-0.4cm}
                                \label{table:numberclients}
                                \begin{center}
                                \begin{tabular}{|c|c|c|c|}
                                \hline
                                
                                Clients&  \thead{GAS Pipeline \\ Target accuracy: 94.7\%}  &  \thead{WUSTL-IIoT \\ Target accuracy: 99.8\%}\\ \hline
                                2 &  551.12 & 1562.53 \\ \hline
                                5& 1655.78 & 5965.07 \\ \hline
                                10& 2483.67 & 10737.15 \\ \hline
                                \end{tabular}
                                \end{center}
                                \vspace{-0.1cm}
                                \end{table}

It is evident from Table \ref{table:numberclients} that the overall training time for BFL increases as the number of clients increases. This is a result of two factors. Firstly, when datasets contain specific samples, dividing the data among a greater number of clients results in each client getting a smaller portion of the overall dataset. As a result, each client must train on a smaller dataset, resulting in slower convergence. Secondly, the increase in the number of clients requires the server to communicate with more clients to produce the aggregated model, which results in the algorithm taking longer to converge to the target accuracy.

\subsection{Efficacy of communication and the Straggler effect of BFL}\label{yy}
Accuracy and convergence speed allow 
to evaluate FL methods' robustness against stragglers and communication efficiency. We consider the same target accuracies that are mentioned earlier when evaluating the convergence speed of each FL method. Moreover, we compare model accuracy based on 5 clients, 10 iterations for the Gas pipeline dataset, and 4 iterations for the WUSTL-IIoT dataset. SFL and AFL  have foundational implementations detailed in \cite{FEDAVG} and \cite{asynchFL}, respectively.
Although FedBuff's accuracy and convergence speed increase with increasing buffer size, choosing the appropriate buffer size is its weakness. Therefore, we assume that the buffer size in FedBuff is set to half the number of clients participating in the heterogeneous environment. Accordingly, for a scenario with five clients, we consider a buffer size of two.

To simulate speed heterogeneity, we assigned 5 random integer values to 5 clients (1$\sim$ 3s to three of them and 5$\sim$10s to two of them). 
A higher random value indicates a slower client, whereas a lower random value indicates a faster client. Additionally, to better demonstrate the effectiveness of the proposed model in scenarios involving heterogeneous datasets, we assume varying data sizes for each client. This means that clients with more data will have longer training times, while those with less data will train more quickly.

To simulate such a scenario, each of the three fast clients is allocated $5\%$ random samples from the training dataset, while the two slow clients are randomly assigned $80\%$ and $90\%$ samples of the training data, respectively.  

\begin{figure}[htbp]
            \centerline{\includegraphics[width=2.40in]{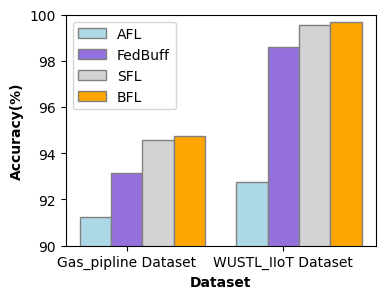}}
            \vspace{-0.2cm}
            \caption{Accuracy comparison of MLP-based algorithm: 5 clients, 10 iterations for Gas pipeline and 4 iterations for WUSTL-IIoT.}
            \label{figure:acccomparison}           
            \end{figure}

The results depicted in Fig. \ref{figure:acccomparison} demonstrate the superiority of BFL in achieving higher attack detection accuracy, compared to baseline FL methods. We observe that BFL outperforms SFL, AFL and FedBuff in terms of accuracy by, respectively, $0.18\%$, $3.53\%$, and $1.63\%$ on Gas pipeline dataset and $0.13\%$, $6.9\%$, and $1.1\%$ on WUSTL-IIoT dataset. This is attributed to BFL's novel weighted average time, which can more effectively engage straggling clients, resulting in better prediction results (3 clients are selected by BFL). As SFL and BFL follow the same synchronous updating strategy, they have the closest prediction performance. AFL achieved the lowest accuracy, which can be explained by the fact that AFL aggregates weights from one client at a time and does not provide an effective way to deal with stragglers. In addition, FedBuff is unable to converge faster because its buffer is rapidly filled with clients that have less data and less accuracy.

            \begin{figure}[htbp]
            \centerline{\includegraphics[width=2.40in]{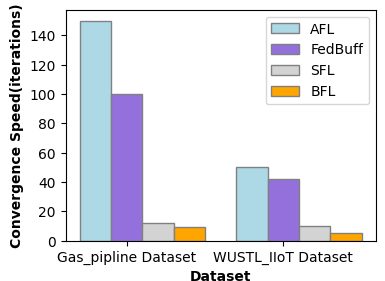}}
            \vspace{-0.2cm}
            \caption{Convergence speed comparison of MLP-based algorithms.}
            \label{figure:converspeed}
            
            \end{figure}

A comparison of the convergence speed results shown in Fig. \ref{figure:converspeed} also reveals a clear difference in performance. On both datasets, BFL converges faster towards the target accuracy than all three other methods. On Gas pipeline dataset, BFL convergences $1.33$, $16.66$, and $11.11$ times faster than SFL, AFL and FedBuff, respectively. Similarly, BFL surpasses SFL, AFL and FedBuff by, respectively, $2$, $10$, and $8.4$ times on WUSTL-IIoT dataset. The faster convergence of BFL is the result of the balance that BFL creates between fast and slow clients, avoiding to wait for the slowest client like in SFL or favor faster clients as in FedBuff. Additionally, the convergence speed result for AFL indicates that AFL methods, where the server simply communicates with all clients, suffer from a severe communication bottleneck.

Hence, considering both attack detection performances and convergence speed, BFL provides the best performance-cost balance.

\section{Conclusion}\label{conc}
\vspace{-0.06cm}

In this paper, we proposed a novel FL framework, named BFL, to detect cyber threats against industrial CPSs. BFL employs a HE-based secure communication protocol that preserves the privacy of model parameters during training. In BFL, the following modules are synthesized cohesively: (1) a buffering strategy to handle stragglers; (2) synchronous updating of the global model to prevent communication bottlenecks; (3) a novel, weighted average time method that the FL server uses to balance between fast and slow clients in a heterogeneous environment. In extensive experiments using two real industrial CPS datasets, BFL achieved the highest prediction accuracy, converged the fastest, and had superiority over state-of-the-art FL schemes. {{In future work, we intend to improve the weighted average time method }}to dynamically adapt the selection of clients at each training iteration based on the changing computation capability of clients and the fluctuating network conditions.

\vspace{-0.2cm}
\section*{Acknowledgment}

This research work is partially supported by the Business Finland 6Bridge 6Core project under Grant No. 8410/31/2022, the Research Council of Finland 6G Flagship program (Grant No. 346208), and the European Union’s Horizon Europe research and innovation programme HORIZON-JU-SNS-2022 under the RIGOUROUS project (Grant No. 101095933). The paper reflects only the authors' views. The Commission is not responsible for any use that may be made of the information it contains.
\vspace{-0.2cm}


\end{document}